\documentclass[aps,prl,reprint]{revtex4-2}
\usepackage[utf8]{inputenc}
\usepackage{textcomp}
\usepackage{amsmath}
\usepackage{amssymb}
\usepackage{graphicx}
\usepackage{color}
\usepackage{dsfont}
\usepackage{bbm}
\usepackage{siunitx}
\usepackage{wrapfig}

\begin{document}

\title{Superfluid Helium Drops Levitated in High Vacuum}
\author{C. D. Brown,$^{1,2}$ Y. Wang,$^3$ M. Namazi,$^{1,2}$ G. I. Harris,$^{1,4}$ M. T. Uysal,$^1$ and J. G. E. Harris$^{1,2,3}$}
\affiliation{$^1$Department of Physics, Yale University, New Haven, CT 06520}
\affiliation{$^2$Yale Quantum Institute, Yale University, New Haven, CT 06520}
\affiliation{$^3$Department of Applied Physics, Yale University, New Haven, CT 06520}
\affiliation{$^4$ARC Centre of Excellence for Engineered Quantum Systems, School of Mathematics and Physics, The University of Queensland, Brisbane, QLD, Australia}

\date{\today}


\begin{abstract}
We demonstrate the trapping of millimeter-scale superfluid Helium drops in high vacuum. The drops are sufficiently isolated that they remain trapped indefinitely, cool by evaporation to $330$ mK, and exhibit mechanical damping that is limited by internal processes. The drops are also shown to host optical whispering gallery modes. The approach described here combines the advantages of multiple techniques, and should offer access to new experimental regimes of cold chemistry, superfluid physics, and optomechanics.

\end{abstract}

\maketitle

Liquid helium drops offer a combination of isolation, low temperature, superfluidity, and experimental access that is unique among condensed matter systems. These features make it possible to address a number of questions in chemistry and physics~\cite{VilesovChapter,Northby2001}. For example, He drops have been used to cool a range of molecular species well below $1$ K, facilitating precision spectroscopy and studies of cold chemical reactions~\cite{Lehman2006,Scoles2001,Vilesov2000,Vilesove1998}. Drops of pure He can be used to explore quantum many-body phenomena such as the microscopic character of superfluidity and the condensate fraction in strongly interacting sytems~\cite{Kwon2000, Paulson2003, Halley1993}. They also offer access to outstanding issues in classical and quantum fluid dynamics, including the interplay of turbulence, vorticity, and topology ~\cite{Vilesov2020,Vilesov2014,Chandrasekhar1965,Scriven1980,Heine2006,Pi2019}. Lastly, He drops that support optical whispering gallery modes (WGMs) have been proposed as a system for exploring macroscopic quantum phenomena~\cite{Harris2017}.

The scientific questions that can be addressed with a He drop depend on the drop's size, temperature, and degree of isolation. These parameters determine whether the drop can become superfluid, host  chemical dopants, and support WGMs. They also set the frequencies and damping rates of the drop's excitations (such as its bulk and surface acoustic modes), and hence the time scale over which these exctitations retain quantum coherence. 

To date, experiments with superfluid drops have followed one of two broad approaches. In the first, liquid He is injected into a vacuum chamber, producing drops with radius $1$ nm $\lesssim R \lesssim 1$  $\mu$m that travel ballistically through the chamber~\cite{Winkler1990,Toennies1994,Toennies1997_1,Toennies2001,Toennies2004,Ernst2015_1,Ernst2015_2}. This approach provides sufficient isolation for the drops to evaporatively cool well below the temperature of the chamber walls (to $T_\mathrm{drop} \approx 380$ mK ). Such drops become superfluid and can host dopants. However their small size limits the range of fluid dynamics they can access, and precludes them from supporting WGMs. Furthermore, they travel at $\sim 300$ m/s, limiting their lifetime (they collide with the chamber wall in $\sim$ ms) and the range of experimental probes that can be applied to them.

The second approach uses traps to achieve much longer interrogation times and larger drops. Stable trapping has been achieved using magnetic~\cite{Maris1996}, optical~\cite{Seidel1995}, and electrical forces~\cite{Niemela1997}, and with $1$ $\mu$m $\lesssim R \lesssim 10$ mm. However, to date trapping has been achieved only in the presence of He vapor that prevents the drop from achieving isolation. Background He vapor has prevented trapped drops from cooling below the temperature of their enclosure, and in most studies has dominated the damping of their motion.

In this paper, we demonstrate stable magnetic trapping of mm-scale superfluid drops in high vacuum. We show that this approach combines the advantages of the ballistic method (isolation and evaporative cooling) with the advantages of trapping (long interrogation times and large drops). We measure the trapped drops' thermal and mechanical properties, and also demonstrate that they support optical WGMs. 


A schematic illustration of the experiment is shown in  Ref.~\cite{SI}. Levitation is provided by a superconducting solenoid housed in the $^4$He bath space of a cryostat. The solenoid is designed so that stable levitation is achieved for $115~\mathrm{A}< I <118~\mathrm{A}$, where $I$ is the current in the solenoid. Varying $I$ within this range translates the levitation point vertically, and can be used to vary the drop shape (i.e., from prolate to oblate)~\cite{Maris1997}. Drops are produced and trapped in a custom-built cell that fits in the cryostat's vacuum space and extends into the magnet's bore. The temperature of the cell walls $T_{\mathrm{cell}}$ is controlled by a liquid $^4$He flow line. Optical access to the trapping region is provided by windows in the cryostat and cell \cite{SI}.

\begin{figure}
\includegraphics[scale=0.9]{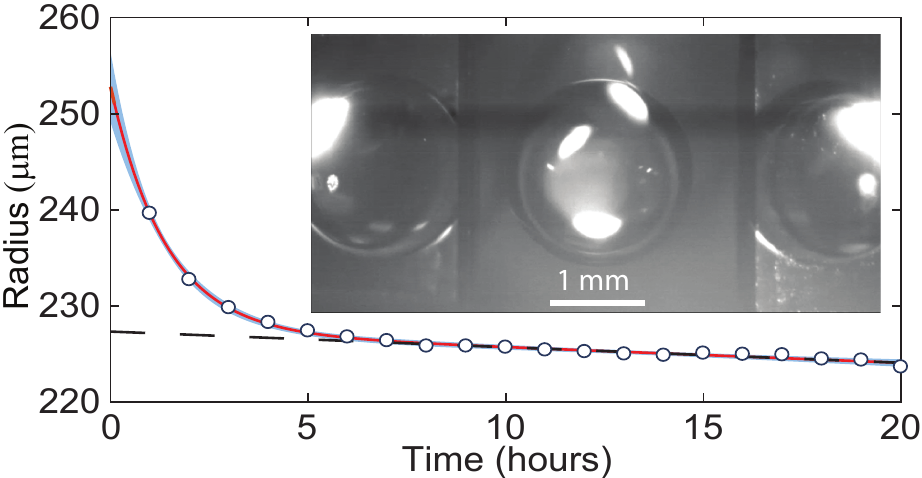}
\caption{A levitated drop. Inset: a $^4$He drop shortly after it has been levitated. The central portion of the image is a direct view of the drop, while the left and right portions are the reflections from two 45$^{\circ}$ mirrors placed near the levitation region. Main figure: The drop radius $R$ (circles) as a function of time. The red curve is a fit to the sum of an exponential and a linear function (the linear portion is the dashed line). The statistical uncertainty in $R$ is $\sim 10$ nm. The blue band shows the systematic uncertainty ($95 \%$ confidence interval).}
\label{fig:newFig2_v4}
\end{figure}

To produce a levitated drop, $I$ is fixed and the cell is cooled by the $^4$He flow line. The cell is then filled with a controlled quantity of $^4$He, which produces a puddle at the bottom of the cell. Next, the cell is opened to a turbomolecular pump (TMP), which causes the puddle to boil aggressively. In the subsequent seconds, a fog of \textmu m-scale droplets aggregates in the levitation region and then coalesces into a single mm-scale drop at the levitation point. The inset of Fig.~\ref{fig:newFig2_v4} shows a levitated drop with $R=1.0$ mm roughly $1$ s after opening the cell to the TMP.

After the drop has been trapped, the TMP continues to evacuate the cell. After roughly five minutes the puddle is completely depleted, and $P_{\mathrm{cell}}$ decreases sufficiently that thermal contact between the drop and the cell walls is broken. The drop's thermal isolation is evidenced by the fact that $R$ appears constant (within the resolution of the imaging system) for several hours.

However, close examination shows that the drop continues to evaporate, albeit very slowly. To measure the slow change in $R$, we use standard image processing techniques~\cite{brown2019optical} to determine the drop's edge in each video frame. This shape is fit to a circle, and the value of $R$ returned by this fit is averaged over 1,200 images (acquired in 60 s) to produce each of the data points shown in Fig.~\ref{fig:newFig2_v4}. This data shows that the evaporation rate  decreases in the first few hours after trapping, and then becomes roughly constant. A linear fit to the last 12 hours of data gives an average evaporation rate $\Dot{R}= (0.44\pm 0.04)~\mbox{\AA}/\mathrm{s}$. According to the model described in Ref.~\cite{BrinkStringari1990}, this corresponds to {$T_{\mathrm{drop}}\approx 330$} mK and a heat load  $\dot{Q} \sim 30$ pW on the drop. As described below, the likely source of this heat is a small amount of residual He vapor in the cell.

The drop's center-of-mass (COM) motion is measured using a laser (DL) with wavelength $\lambda = 1,550$ nm which passes though the drop so that it is refracted by an angle that depends on the drop’s position. This deflection is measured using a photodiode \cite{SI}.

\begin{figure}
\centering
\includegraphics[scale=1]{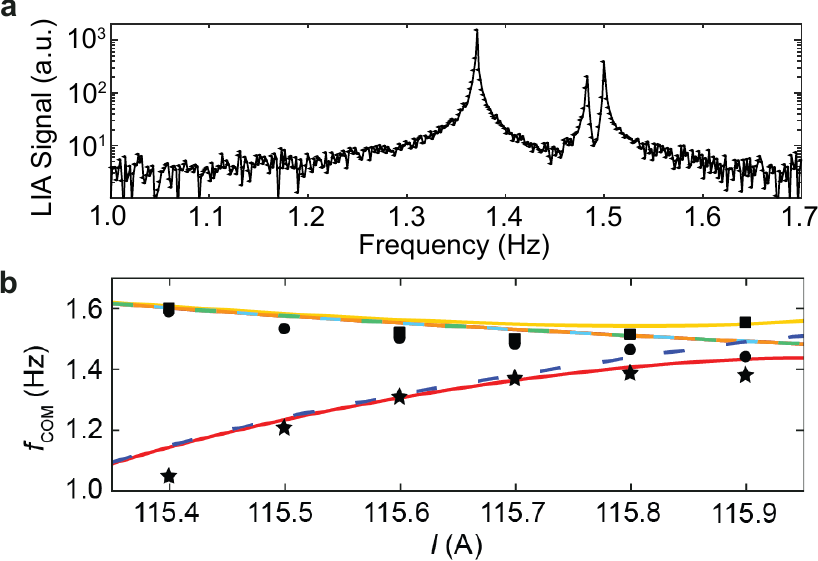}
\caption{Center-of-mass (COM) motion of a levitated drop. (a) The power spectral density of the COM motion for $I = 115.7$ A. (b) The frequencies of the normal modes versus the magnet current. Black markers: frequencies determined by fitting the data in (a). Dashed lines: the calculated radial (light blue and light green) and axial (dark blue) frequencies assuming the magnets axis is parallel to gravity. Solid lines (red, orange, yellow): the best fit of the data for a magnetic trap that is tilted with respect to gravity. }
\label{fig:newFig3_v4_test}
\end{figure}

Fig.~\ref{fig:newFig3_v4_test}a shows a typical spectrum of the COM motion. No deliberate drive was applied to the drop; the observed motion is the drop's steady-state response to vibrations in the cryostat.  For each value of $I$, the data show peaks corresponding to the three normal modes of motion in the trap. The resonant frequencies $f_{\mathrm{COM}}$ of these modes are shown as a function of $I$ in Fig.~\ref{fig:newFig3_v4_test}b. The dashed lines are the frequencies calculated (without free parameters) for a trapping field whose symmetry axis is colinear with gravity. In this model, the radial and axial frequencies are $ \omega_r^2=(-\chi / \mu_0 \rho)((\tfrac{1}{2}\partial_z B_z)^2-\tfrac{1}{2}B_z \partial_{z z} B_z)$ 
and 
$\omega_z^2=(-\chi / \mu_0 \rho)((\partial_z B_z)^2+B_z \partial_{z z} B_z)$ respectively, where $\rho=145~\mathrm{kg}/\mathrm{m}^3$  and $\chi=-9.85\times10^{-10}$ are the density and the volume diamagnetic susceptibility of $^4$He. The magnetic field and its derivatives are evaluated at the levitation point \cite{brown2019optical} (these quantities are known from the magnet design). 

While this model reproduces the qualitative features in the three $f_{\mathrm{COM}}(I)$, it does not capture their behavior near the predicted degeneracy at $I=115.9$ A. The solid lines in Fig.~\ref{fig:newFig3_v4_test}b show a fit to a model that incorporates a relative angle $\theta$ between gravity and the trap's symmetry axis~\cite{brown2019optical}. Using $\theta$ as a fitting parameter returns $\theta= (0.27\pm0.11)^{\circ}$. This misalignment  may result from an actual tilt of the cryostat, or from deformation of the trapping fields due to the magnetic response of the cell materials.


\begin{figure} 
\centering
\includegraphics[scale=0.55]{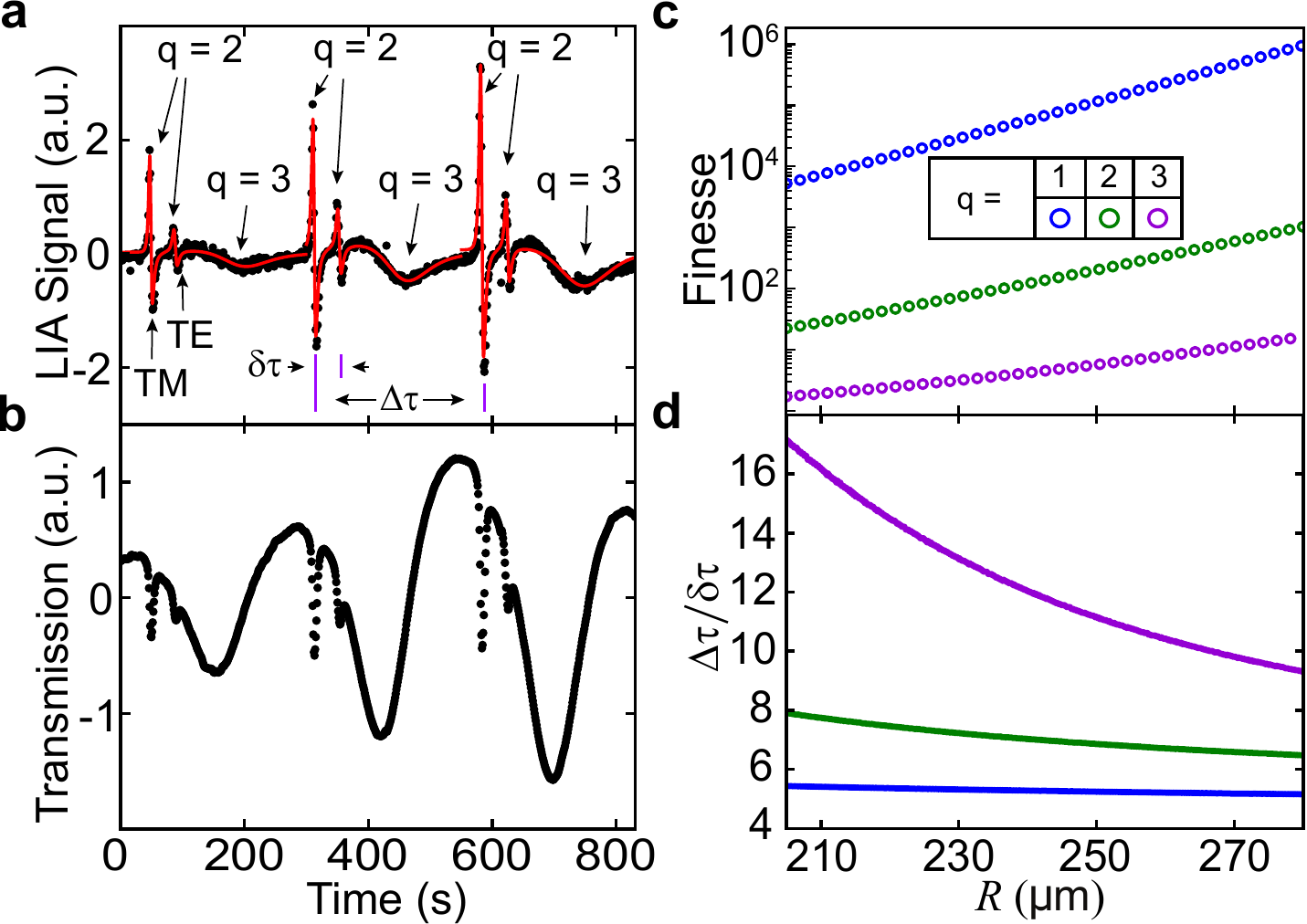}
\caption{Optical WGMs. (a) The lock-in signal produced by optical transmission through a superfluid drop with $R=240\pm 1~\mu\mathrm{m}$. (b) The integral of the data in (a). (c) The calculated finesse for TE WGMs with $q \in \{1,2,3\}$. The values for TM WGMs are nearly identical. (d) The calculated splitting between TE and TM modes, $\Delta \tau/\delta t$, with $q \in \{1,2,3\}$.}
\label{fig:newFig4_v6}
\end{figure}

The drops levitated here are nearly spherical, with index of refraction  $n_{\mathrm{He}}=1.028$ for visible and near-infrared wavelengths, and vanishingly small absorption (predicted to be $\sim 10^{-9}$ m$^{-1}$ for $T_{\mathrm{drop}} = 330$ mK ~\cite{seidel2002rayleigh,Seidel1995}). As a result they are expected to host optical WGMs whose finesse increases rapidly with $R$ for $R > 0.1$ mm~\cite{Harris2017}.



To characterize these WGMs, we use the setup shown in Ref.~\cite{SI}. The DL is focused at the center of the drop and its intensity is modulated at a frequency close to the resonance of the drop's $\ell_{\mathrm{cap}} = 2$ capillary mode (described below). The optical dipole force exerted by the DL beam excites this capillary mode, which effectively modulates $R$ (more precisely, the drop's circumference in the plane of the WGMs is modulated). At the same time, an intensity-stabilized HeNe laser ($\lambda = 633$ nm) is focused at the drop's edge, and its transmission is recorded using a lock-in amplifier (LIA). In addition to the modulation produced by the drop's capillary mode, the drop's evaporation causes $R$ to slowly decrease with time. As a result, the LIA signal is approximately proportional to the derivative of the drop's transmission with respect to $R$.

Fig.~\ref{fig:newFig4_v6}a shows a typical record from the LIA for a drop trapped with $I = 116$ A. Analysis of video images taken during these measurements gives $R=240\pm 1$ \textmu m.  Fig. \ref{fig:newFig4_v6}b shows the same data integrated with respect to time, giving a signal proportional to the optical transmission through the drop. The data show a pattern of features that repeats with a period $\Delta \tau \sim 300$ s. Each feature corresponds to a WGM being tuned through resonance with the HeNe by the drop's evaporation. Each repetition of the pattern corresponds to the drop's circumference changing by $\lambda_{\mathrm{HeNe}}/n_{\mathrm{He}}$ (equivalent to the WGM's angular index $\ell \approx 2,380$ changing by $1$), which tunes the cavity through one free spectral range (FSR). 

Within each of the three FSRs shown in Fig.~\ref{fig:newFig4_v6}a, the data is fit to the sum of three (once-differentiated) Lorentzians, with each Lorentzian's center position, linewidth, and amplitude used as fit parameters. The result is the red curve in Fig.~\ref{fig:newFig4_v6}a. These fits give the finesse $\mathcal{F} = 36 \pm 2$ for the largest feature, $\mathcal{F} = 30 \pm 3$ for the middle feature, and $\mathcal{F} = 1.9 \pm 0.1$ for the broadest feature (these  values are the averages over the three FSRs shown in Fig.~\ref{fig:newFig4_v6}a).

To determine the identities of these modes, Fig.~\ref{fig:newFig4_v6}c shows the calculated $\mathcal{F}$ for WGMs in a sphere with index of refraction $1.028$, as a function of the sphere's radius~\cite{Oraevsky}. Results are shown for both TE and TM polarizations, and for values of the WGM's radial index $q \in \{1,2,3\}$ (where $q-1$ gives the number of a radial electric field nodes within the drop). Fig.~\ref{fig:newFig4_v6}d shows the calculated splitting between TE and TM modes (having all other mode indices equal). These plots indicate that the broadest feature in each FSR corresponds to $q=3$ modes (their linewidth is too large to resolve the  TE and TM modes separately), and that the two narrower features correspond to TE and TM modes with $q=2$. 

The measured linewidths of these $q=2$ modes are roughly three times greater than in the calculation shown in Fig.~\ref{fig:newFig4_v6}c. This is consistent with the ellipticity ($\epsilon \sim 10^{-5}$) expected for this value of $R$ and $I$~\cite{Maris1997}. Specifically, $\epsilon$ splits the degeneracy over the WGM's azimuthal index $m$ into resonances whose splittings (i.e. between modes with $m$ differing by unity) are all much smaller than the expected WGM linewidth. As a result, they should form an unresolved band whose width would correspond to an apparent finesse $\mathcal{F}_{\epsilon} = 46$ for the $q=2$ modes.

The fit in Fig.~\ref{fig:newFig4_v6}a also gives the ratio between the FSR and the splitting between the TE and TM $q=2$ modes as $6.6 \pm 0.1$. This is in good agreement with the calculated value of $6.9$ (Fig.~\ref{fig:newFig4_v6}d).

We did not observe the $q=1$ WGMs, whose finesse is expected to be $\sim 10^4$. This is likely because of poor mode-matching between these modes and the HeNe beam, and because the drop's evaporation tuned these modes through resonance too quickly to be recorded with our data sampling rate ($1$ Hz).

\begin{figure} 
\centering
\includegraphics[scale=1.1]{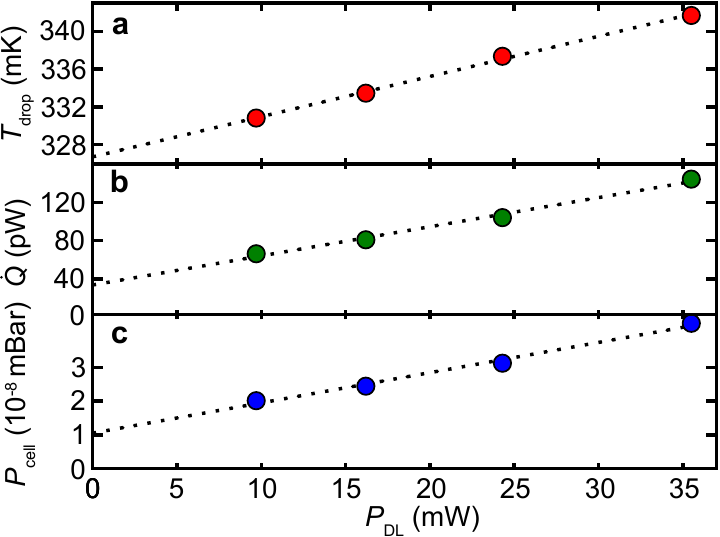}
\caption{The drop's thermal properties. (a) The drop temperature. (b) The heat load on the drop. (c) The background pressure in the cell. These quantities are obtained from measurements of $\dot{R}$, and are plotted as a function of the power of the laser incident on the drop. The dashed lines are linear fits.}
\label{fig:newFig5_v4}
\end{figure}

Since the passage of each FSR corresponds to the drop circumference changing by $\lambda_{\mathrm{HeNe}}/n_{\mathrm{He}}$, we can use $\Delta \tau$ as a measurement of the drop's evaporation rate $\dot{R}=\lambda_{\mathrm{HeNe}} / 2\pi n_{\mathrm{He}} \Delta \tau$. The evaporation model given in Refs.~\cite{BrinkStringari1990,Harris2017} can then be used to infer $T_{\mathrm{drop}}$ and $\dot{Q}$ from $\dot{R}$. This approach is illustrated in Fig.~\ref{fig:newFig5_v4}, which shows data for a drop with $R=207.5\pm 1~\mu \mathrm{m}$ (as determined by image analysis). The optical transmission through this drop (not shown) has features similar to those in Fig.~\ref{fig:newFig4_v6}a, which are fit to determine $\Delta \tau$. Figs.~\ref{fig:newFig5_v4}a,b show $T_{\mathrm{drop}}$ and $\dot{Q}$ inferred in this manner as a function of $P_{\mathrm{DL}}$, the power of the DL incident on the drop. The data are consistent with a heat load proportional to $P_{\mathrm{DL}}$, along with a background heat load $\sim 35$ pW. While the former contribution could reflect absorptive heating of the drop by the DL, the coefficient of proportionality ($3 \times 10^{-9}$) is roughly three orders of magnitude greater than expected~\cite{seidel2002rayleigh,Seidel1995}. If, instead, the observed heatload is attributed to He gas in the cell (assumed to be at the temperature of the cell walls), the corresponding pressure $P_{\mathrm{cell}}$ is shown in Fig.~\ref{fig:newFig5_v4}c. We attribute the increase in $P_{\mathrm{cell}}$ with increasing $P_{\mathrm{DL}}$ to the absorption of laser light by various objects in the cell.

Vibrations of the drop for which the restoring force is dominated by surface tension are known as capillary modes. These modes' oscillation frequencies are given by
\begin{equation}
\label{eqn:capillaryfreqs}
f_{\ell_{\mathrm{cap}}}=\sqrt{\ell_{\mathrm{cap}}(\ell_{\mathrm{cap}}-1)(\ell_{\mathrm{cap}}+2)\sigma/4 \pi^2 \rho R^3}
\end{equation}
where $\ell_{\mathrm{cap}} \in \{2,3,4,...\}$ and $\sigma=3.75\times10^{-4}~\mathrm{J}/\mathrm{m}^2$ is the surface tension of superfluid liquid $^4$He~\cite{Marris2002}. To drive these modes, the DL is focused at the drop's center and its intensity is modulated at frequency $f_{\mathrm{drive}}$. The modes' response is monitored by recording the transmission of the HeNe beam through the drop. This beam's position is chosen to avoid the optical WGMs, so its transmission is modulated because the capillary modes deflect the beam. 

\begin{figure} [h]
\centering
\includegraphics[scale=.85]{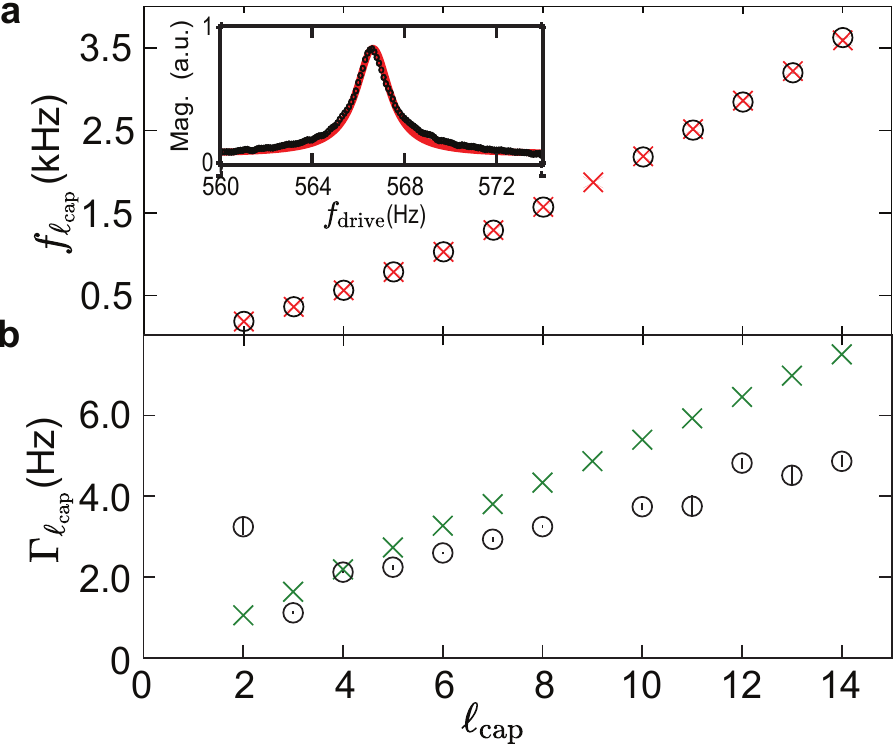}
\caption{Capillary mode resonances. (a) The measured (circles) and expected (crosses) resonance frequencies. Inset: the resonance of the $\ell_{\mathrm{vib}}=4$ mode (black), along with the best fit (red). (b) The measured (circles) and expected (crosses) damping rates.}
\label{fig:newFig6_v6}
\end{figure}

Fig.~\ref{fig:newFig6_v6} shows the frequencies and linewidths of the first several resonances measured in a drop with $R=246 \pm 0.7~\mu\mathrm{m}$ and $T_{\mathrm{drop}}\approx 330$ mK. The frequencies and linewidths are determined by fitting each resonance. Assuming that each resonance corresponds to a distinct value of $\ell_{\mathrm{cap}}$ (except for $\ell_{\mathrm{cap}} = 9$, which did not produce a measurable signal) the resonance frequencies are found to agree with Eq. \ref{eqn:capillaryfreqs} to better than $1 \%$.

These modes' linewidths $\Gamma_{\ell_\mathrm{cap}}$ are shown in Fig. \ref{fig:newFig6_v6}b, along with the values expected from the damping of capillary modes by inelastic scattering of thermal phonons from the drop's surface: \cite{Williams1996}

\begin{equation}
\label{eqn:surfModeDecayRate}
\frac{\Gamma_{\ell_{\mathrm{cap}}}}{2\pi}=\frac{\pi^{2} \hbar \mathcal{K}}{60 \rho_{0}}\left(\frac{k_{\mathrm{B}} T}{\hbar u_{\mathrm{c}}}\right)^{4}
\end{equation}
where $\mathcal{K}=(\ell_{\mathrm{cap}}(\ell_{\mathrm{cap}}-1)(\ell_{\mathrm{cap}}+2))^{1/3}/R$ and $u_{\mathrm{c}}=238~\mathrm{m}/\mathrm{s}$ is the speed of sound in liquid $^4$He. While this prediction shows qualitative agreement with the data, we note two  discrepancies. The first is in the average slope of $\Gamma_{\ell_\mathrm{cap}}$ vs. $\ell_\mathrm{cap}$. This slope is predicted to be $\propto T_{\mathrm{drop}}^{4}$, and would agree with the observed slope if one were to take $T_{\mathrm{drop}} = 310$ mK. However this would correspond to an evaporation rate $\sim 4 \times$ smaller than observed. The second discrepancy is in the damping rates for $\ell_{\mathrm{cap}} = 2$ and $\ell_{\mathrm{cap}} = 3$, which depart from the simple trend predicted by Eq.~\ref{eqn:surfModeDecayRate}.

Both discrepancies may have their origin in the fact that Eq. \ref{eqn:surfModeDecayRate} is derived under the  assumption that phonons which are inelastically scattered by the surface fully thermalize before scattering from the surface again. However the mean free path of phonons $\Lambda \propto T^{-4}$, with $\Lambda = 4.5$ mm for $T = 330$ mK~\cite{Maris_RMP77}. Furthermore the phonon thermalization time $\Lambda/u_{\mathrm{c}} \approx 16~\mu \mathrm{s} \ll f_{\ell_{\mathrm{cap}}}^{-1}$ for $2\leq \ell_{\mathrm{cap}} \leq 14$. Thus, a thermal phonon in the drops studied here will scatter many times from an effectively stationary drop surface. The damping of capillary modes in this regime has not been calculated.

In conclusion, we have shown that mm-scale drops of superfluid $^4$He can be magnetically levitated in high vacuum indefinitely, and that their thermal, optical, and mechanical properties are consistent with expectations. The combination of isolation, evaporative cooling, long measurement time, and large drop size that is demonstrated here opens a number of new avenues for exploration with superfluid He drops. These include more precise measurements of the spectra and chemical reactions of cold molecules, studies of the mechanical damping produced by non-Markovian baths, and the study of the onset and decay of superfluid turbulence in a wall-free system. In addition, we expect improvements in the experimental cell to further reduce the density of background He gas, resulting in lower drop temperature and correspondingly lower mechanical damping and evaporation, while the use of \textit{in situ} mode-matching optics and improved data acquisition should allow access to the drops' high-finesse $q=1$ WGMs. The realization of such WGMs in objects whose stiffness is set only by the weak surface tension of liquid He may provide access to new regimes of quantum optomechanics~\cite{Harris2017}.

We acknowledge helpful discussions with Andrea Aiello,  Rosario Bernardo, Vincent Bernardo, Lilian Childress, Leonid Glazman, Aaron Hillman, David Johnson, Florian Marquardt, Craig Miller, and Yogesh Patil. We acknowledge support from W. M. Keck Foundation Grant No. DT121914, AFOSR Grant No. FA9550-15-1-0270, DARPA Grant No. W911NF-14-1-0354, ARO Grant No. W911NF-13-1-0104, NSF Grant No. 1707703, ONR Grant N00014-18-1-2409, and Vannevar Bush Faculty Fellowship N00014-20-1-2628. This project was made possible through the support of a grant from the John Templeton Foundation. This material is based upon work supported by the NSF Graduate Research Fellowship under Grant No. DGE-1122492 and by a Ford Foundation Dissertation Fellowship.


\bibliographystyle{apsrev4-1}
\bibliography{HeDropBib2}

\newpage
\onecolumngrid
\section{Supplemental Information for ``Superfluid Helium Drops Levitated in High Vacuum''}

\maketitle

\section*{Introduction}
In this supplementary material we provide details of our experimental setup, and provide additional data on the drops' vibrational modes. Section~\ref{section:ExpSetup} describes the levitation cryostat and the laser system used to probe levitated drops. Section \ref{section:SurfModes_VS_R} describes measurements of the resonant frequency and decay rates of drops' surface modes as a function of the drops' radii.


\section{Experimental Setup}
\label{section:ExpSetup}

\begin{figure}[h] 
\centering
\includegraphics[scale=.9]{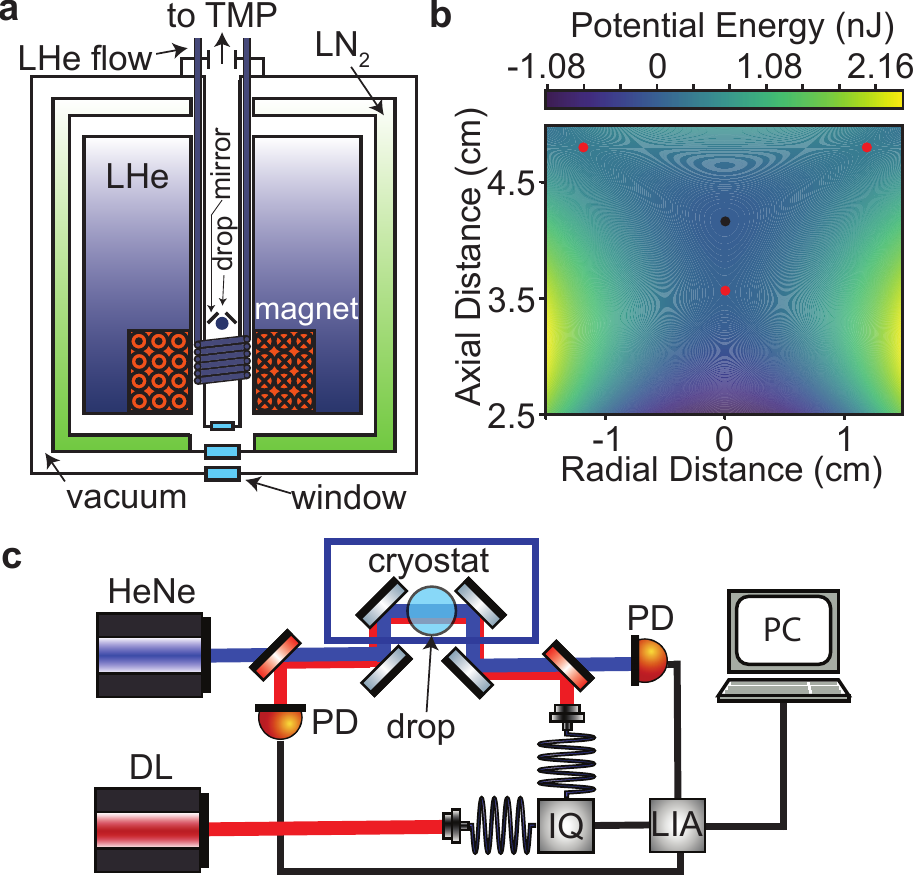}%
\caption{Experimental schematic. (a) Cross-sectional view of the cryostat and the experimental cell. (b) The potential energy of a drop (with $R = 1$ mm), showing a stable equilibrium point (black circle) and saddle points (red circles). The axes are the axial and radial distance from the solenoid's center. (c) The measurement apparatus, showing the two lasers, optical modulator (IQ), photodiodes (PD), and lock-in amplifier (LIA)}
\label{fig:newFig1_v5}
\end{figure}

\hspace{6mm}Fig.~\ref{fig:newFig1_v5}a provides an illustration of the non-uniform superconducting solenoid housed in the $^4$He bath space of the cryostat, which we use to levitate the drops. We produce and trap drops in a custom-built cell that fits in the cryostat's vacuum space and extends through the magnet's bore.

As described elsewhere, levitation occurs when $B_{z}\partial_{z} B_{z}=\mu_0 \rho g/ \lvert \chi\rvert$, with this levitation point being stable when $\partial_{ii}B^2>0$ for all $i  \in \{x,y,z\}$~\cite{Maris1996}. Here, $z$ is the axial coordinate, $B_{z}$ is the axial magnetic field component, $\mu_0$ is the permeability of free space, $\rho=145~\mathrm{kg}/\mathrm{m}^3$ is the density of liquid $^4$He, $g$ is the gravitational acceleration and $\chi=- 9.85\times 10^{-10}$ is the volume diamagnetic susceptibility of $^4$He. Fig.~\ref{fig:newFig1_v5}b illustrates the magneto-gravitational potential energy of a levitated drop when the magnet is driven with current $I=116$ A.

Fig.~\ref{fig:newFig1_v5}c illustrates the general scheme used to apply optical forces to drops, probe the drops' optical WGMs and to measure the drops' vibrational and optical response. The HeNe laser is used to detect the drop's vibrations or optical WGMs, while the DL is intensity-modulated to apply optical forces to a drop. Both the vibrational and optical response of the drops are inferred from the PD photocurrent using the LIA.

\newpage

\section{Capillary Mode Dependence on Drop Radius}
\label{section:SurfModes_VS_R}

\begin{figure} 
\centering
\includegraphics[scale=.75]{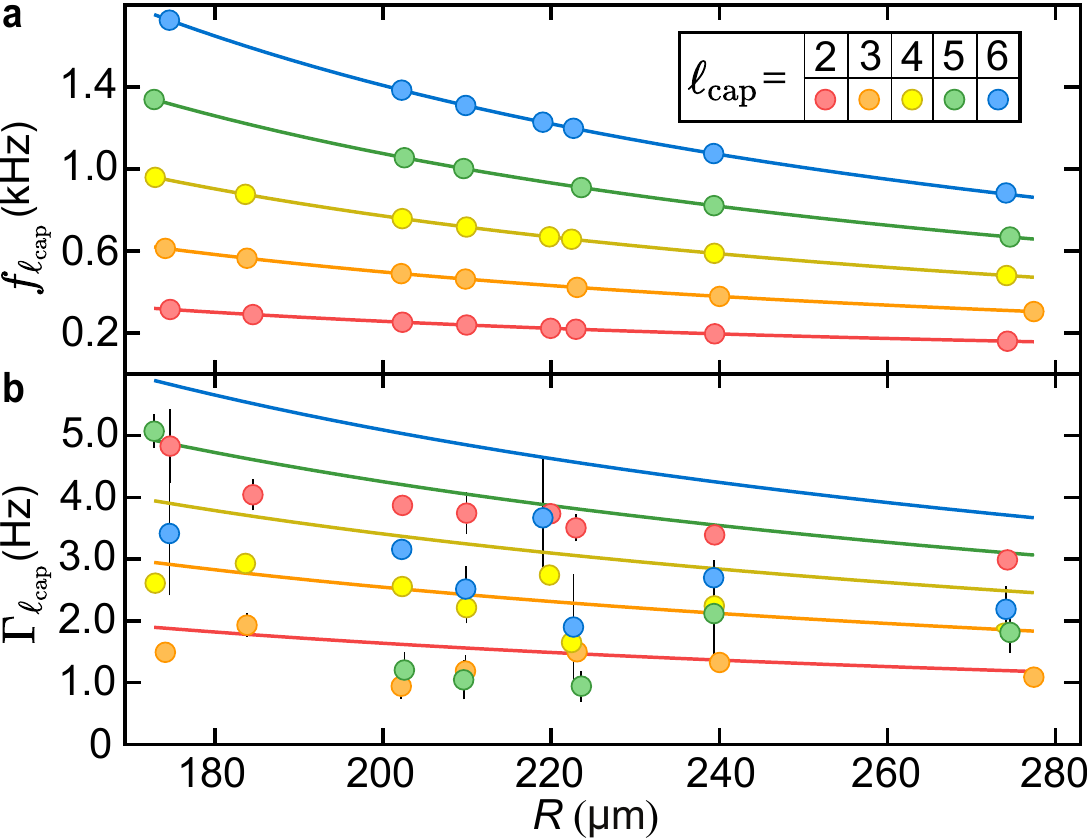}
\caption{The capillary modes' resonant frequencies (a) and damping rates (b) as a function of $R$. The circles are the data and the dotted lines are the expected values assuming  $T_{\mathrm{drop}}=350~\mathrm{mK}$.}
\label{fig:newFig7_v4}
\end{figure}

Fig.~\ref{fig:newFig7_v4} shows the capillary modes' resonant frequencies (Fig.~\ref{fig:newFig7_v4}a) and linewidths (Fig.~\ref{fig:newFig7_v4}b) as a function of the drop radius. As in Fig. 5 of the main text, the resonant frequencies show excellent agreement with Eq. 1 of the main text, while the linewidths show only qualitative agreement with Eq. 2 of the main text.

\end{document}